\documentclass[%
 reprint,
 amsmath,
 amssymb,
 aps,
]{revtex4-1}
	\usepackage{graphicx}
	\usepackage{xcolor}
	\usepackage{dcolumn}
	\usepackage{bm}
	\usepackage{hyperref}
	
	\newtheorem{definition}{Definition}

\begin{document}
\title{Generalised asymptotic classes for additive and non-additive entropies}
\author{Nana Cabo Bizet}
\email{nana@fisica.ugto.mx}
\author{Jes\'us Fuentes}
\email{j.fuentesaguilar@ugto.mx}
\author{Octavio Obreg\'on}
\email{octavio@fisica.ugto.mx}

\affiliation{\small \em Departamento de F\'isica, Divisi\'on de Ciencias e Ingenier\'ias, Universidad de Guanajuato\\
Loma del Bosque 103, Le\'on 37150, Guanajuato, M\'exico.}
\date{\today}
\begin{abstract}

In the field of non-extensive statistical mechanics it is common to focus more attention on the family of parameter-dependent entropies rather than on those strictly depending on the probability in which there is no need to adjust a specific parameter. In particular, there exist two non-parametric entropy measures, $S_\pm$, that resemble the Boltzmann--Gibbs (BG) entropy, $S_B$, in the thermodynamic limit, whereas a difference between them arises in case that the statistical system possesses a small number of microstates. The difference, although slight, accounts for meaningful physical consequences such as effective forces and inner interactions among constituents. Yet, in this letter we are to report some of the analytical attributes associated to entropies $S_\pm$ via the formulation introduced by Hanel \& Thurner. These two functionals allow to construct a generalised classification of entropy measures in terms of their defining equivalence classes, which are determined by a pair of scaling exponents $(c,d)$. As a result, it has been identified that $S_\pm$ and $S_B$ belong to the same asymptotic, equivalence class. The latter is an interesting fact since it does not occur for non-logarithmic, parameter-dependent entropies. Following this scheme, we also briefly discuss the features of other entropy measures such as Tsallis, Sharma--Mittal and R\'enyi in the asymptotic limit.

\begin{description}
\item[PACS numbers]
02.50.Ga; 05.20.-y; 05.70.Ln
\item[keywords]
generalised entropies, non-extensive entropies, asymptotic classification
\end{description}
\end{abstract}
\maketitle
%
%
%
%
\section{Introduction}
\label{intro}

The concept of entropy is fundamental in understanding the large-scale behaviour of either deterministic dynamical statistical systems or complex systems on account of their intrinsic features. In the orthodox statistical mechanics, for example, the Boltzmann-Gibbs statistics are enough to describe with extraordinary accuracy weakly interacting systems and even long-range correlated systems \cite{pennini} in terms of an additive (extensive) entropy measure.

In general, however, it is not the same for the case of complex (highly correlated) systems. Hence, a macroscopical description relying on the basis of BG entropy fails for the reason that additivity is usually not preserved. To circumvent these anomaly, one can attend the description of correlated systems in the light of non-additive (non-extensive) generalised entropies.


An ample number of generalised entropies has been reported in the literature, particularly the families of non-extensive, parameter-dependent entropies have been studied so widely to describe a variety of complex systems \cite{Tsallis2009,PRLWilk}. Nonetheless, in this letter we are mainly interested in the analysis of another family of non-extensive entropies, here named $S_{\pm}$ (or $p$-entropies $\pm$), originally introduced in \cite{obregon10} under the umbrella of super-statistics. We remark that $S_\pm$ do not depend on any free parameters (typically selected in accordance with the problem of interest) but only on the probability, with the special feature that they could provide a modest description for physical systems with a few number of microstates $\Omega$, whereas they completely resemble BG in the thermodynamical limit, as shown in \cite{ogv,gilvi,CaboObregon18}.


Furthermore, these distinctive attributes granted to $S_\pm$, have led us to derive an {entropic form in terms of a pair of scaling exponents $(c,d)$ by applying the formalism introduced in \cite{hanel2011}}. Our results suggest that a wide family of entropies, either parameter-dependent or not, including the family $S_\pm$, can be generated from this new entropic form.

{Our discussion begins by reviewing the asymptotic analysis originally introduced in \cite{hanel2011} --- a formalism that establishes a classification for a number of generalised entropies. Then we study the non-extensive, probability-dependent entropies $S_\pm$ on which our proposal is based. In this regard, we introduce a general entropic form that enables a wide family of entropies to be derivable, such as $S_\pm$. After that, to serve as comparison with $S_\pm$ statistics, we discuss further non-extensive entropies such as Sharma-Mittal, R\'enyi and Tsallis as well as their properties in the asymptotic limit. Finally, we present our conclusions.}

%
%
%
%
\section{Asymptotic analysis}

In what follows we consider generalised entropies and their description in terms of their asymptotic behaviour with respect
to BG entropy. We present the $S_\pm$ entropies, which will be useful in describing asymptotically equivalence classes. We also
demonstrate the convexity of $S_\pm$.

Our study begins by considering generalised entropies of the form \begin{equation}
\label{entropy}
S(P)=\sum_n^\Omega G(p_n),
\end{equation}
for a set $P$ of probabilities $p_n$ defined on it. The functional $G(p)$ is such that the entropic forms defined in that way satisfy at least the three first Shannon--Khinchin (SK) axioms \cite{khinchin}, to wit: {\it continuity} (SK1), {\it maximality} (SK2) and {\it expandability} (SK3). Inasmuch as the fourth axiom, {\it additivity} (SK4), is uniquely fulfilled by the well known Boltzmann entropy functional $S_B\equiv-\sum_n^\Omega p_n\log p_n$, besides R\'enyi entropy, which is also an additive entropy measure, although it is not of the general form \eqref{entropy}.

Either weakly-interacting or non-interacting statistical systems are successfully described by Boltzmann entropy. Only for these systems it is true that {\it additivity} and {\it extensiveness} are both equivalent. This becomes, however, false in case of correlated statistical systems beyond the equilibrium, for which we have that the axiom SK4 is not entirely satisfied. Nonetheless, as will be shown below there exists the possibility to establish a natural connection between correlated and uncorrelated statistical systems by means of the generalised family of entropies
\begin{equation}
\label{spm}
S_+\equiv\sum_n^\Omega\left(1 - p_n^{p_n}\right), \quad \text{and} \quad  S_-\equiv\sum_n^\Omega\left(p_n^{-p_n}-1\right).
\end{equation}
They were firstly introduced in \cite{obregon10} within the super-statistics framework \cite{beck} for the case of a $\Gamma$-distribution characterising a global fluctuation of temperatures. To be more precise, these entropies describe systems possessing a distribution of cells in local equilibrium, over a large time scale but deviating slightly from the global equilibrium. Eventually, it was found that the same entropies arise from a generalisation of the replica trick \cite{CaboObregon18}, a technique introduced in spin glasses. Also, in recent studies, it has been discussed their relevance for physical systems with a few number of micro-states \cite{ogv,gilvi,obregon15}. For example, in molecular dynamics the entropies $S_\pm$ do render an effective repulsive interaction between the constituents \cite{gilvi}, thus feeling the presence of tiny inner forces that otherwise, in a large-scale layout, would be negligible. Furthermore, the quantum aspects as well as the direct implications in the statistics for a Bose-Einstein condensate have been explored in \cite{Obregon18}. We depart from the physical necessity of exploring micro-systems and their associated thermodynamics, and our analysis constitutes an attempt to delineate a route for that purpose.


The entropies in \eqref{spm} do satisfy SK1-SK3 for a system with  $\lesssim 400$ micro-states, whose constituents experience certain interaction with each others, see Fig. \ref{f:splus}. Although, it is interesting to note that in the thermodynamical limit such correlation becomes negligible and SK4 is satisfied as well. In other words, $S_B$ and $S_\pm$ belong to the same asymptotic equivalence class. To see this, we adopt the approach given in \cite{hanel2011} to introduce the following definition.

\begin{definition} Any generalised entropy satisfying the axioms SK1-SK3 is thoroughly characterised by a pair of scaling exponents $(c,d)$ pertaining respectively to the asymptotic laws
\begin{equation}
\label{law1}
\lim_{\Omega\to\infty}\lambda\frac{G\left(\frac{1}{\lambda\Omega}\right)}{G\left(\frac{1}{\Omega}\right)}=\lambda^{1-c}, \quad \lambda \in \mathbb{R},
\end{equation}
with $0<c\leq1$, and
\begin{equation}
\label{law2}
\lim_{\Omega\to\infty}\frac{G\left(\frac{1}{\Omega^{1+a}}\right)}{G\left(\frac{1}{\Omega}\right)}\Omega^{a(c-1)}=(1+a)^d, \quad a\in\mathbb{R}.
\end{equation}
\end{definition}

Taking into consideration the laws \eqref{law1} and \eqref{law2} it is easy to verify that entropies $S_B$ and $S_\pm$ are characterised by the same pair $(c,d)=(1,1)$. In particular, since $c=1$ then to fulfil SK2 it is necessary that $d\geq0$, which is already satisfied, implying that $G$ is a convex function. For classical entropy $S_B$ this is a well known fact and we are to show that this property is also satisfied by $S_\pm$. One can demonstrate the convexity by attending to the condition $tS(p)+(1-t)S(p')\geq S(tp+(1-t)p')$ with $0\leq t\leq1$. Beginning with $S_+$ we get $(1-t)\sum p_n'^{p_n'}- t\sum p_n^{p_n}\geq \delta$ such that $\delta\geq-\sum (p_n'+t(p_n-p_n'))^{(p_n'+t(p_n-p_n'))}$, since the inequalities must be fulfilled for every $p_n$ and $p_n'$ one is able to note that $(1-t)p_n'^{p_n'}>tp_n^{p_n}$ and $\delta\leq0$; therefore $\delta\leq(1-t)p_n'^{p_n'}- tp_n^{p_n}\leq0$, and consequently $S_+$ is convex. Similarly, for $S_-$, we get $tp_n^{-p_n}+(1-t)p_n'^{-p_n'}\geq(t'p_n+(1-t')p_n')^{(-t'p_n-(1-t')p_n')}$, due to the right-hand member attains zero at $t'=p_n'/(p_n'-p_n)$ whereas the left-hand member at $t=-p_n^{p_n}/(p_n'^{p_n'}-p_n^{p_n})$, that is $t'>t$, hence $S_-$ is convex.

%
%
%
%
\section{Generalised entropic forms}

Inspired by the proposal introduced by Hanel \& Thurner \cite{hanel2011}, below we are to provide a general characterisation of the non-extensive entropies \eqref{entropy} in terms of a generalised bi-parametric entropic form on the basis of $S_{\pm}$ statistics. The proposal given the authors assume the violation of the additivity axiom, finding a wide class of generalised entropies albeit rather based on $S_B$ and classified by the exponents $(c,d)$. In the present work we depart from entropies $S_{\pm}$ to encounter also a wide class of entropy measures.

For that we introduce a definition on account of the functional generators
\begin{equation}
\label{generators}
\gamma_+(x)\equiv 1-x^{x}, \quad \text{and} \quad \gamma_-(x)\equiv x^{-x}-1. 
\end{equation}
Finally we give a further discussion on the inherent features of $S_\pm$ and their relation to the class (1,1).

\begin{definition}
\label{l1}
For any pair of scaling exponents $(c,d)$, there is a factor $\sigma=1/[1+(d-1)c]$, and a characteristic function $\phi(x)$, such that the functional generators $\gamma_\pm(x)$ in \eqref{generators} define the set of universal entropic forms
{
\begin{equation}
\label{gen}
G_\pm(x;c,d) \equiv \sigma\left(\phi(x)-\Gamma[d+1,\alpha_\pm(x;c)] e^r\right),
\end{equation}
where $\alpha_\pm(x;c)=1-cW(-\gamma_\pm(x))$, $r>0$, $\Gamma[\cdot,\cdot]$ is the incomplete gamma function and $W(x)$ is the product logarithm (or Lambert function)}.
\end{definition}

The characteristic function, $\phi(x)$, in Definition \ref{l1} is  entirely defined by each entropic form $G_\pm$ and the related scaling exponents $(c,d)$. For instance, the entropy $S_+$, as defined by \eqref{spm}, belongs to the class $(c,d)=(1,1)$, hence $\sigma=1$ and $\alpha=1-W(x^x-1)$, we also select $r=1$. Then its corresponding characteristic function is given by 
\begin{equation*}
\phi_+(x)=2\left(1-x^x+\exp\left[W\left(x^x-1\right)\right]\right), 
\end{equation*}
(we have attached a subscript to distinguish it from the other case) then, {by reducing the incomplete gamma function in \eqref{gen}}, one gets $G_+(x,1,1)=1-x^x$, from which the entropy $S_+$ is immediately recovered. An analogous procedure takes place for the entropy $S_-$, whose characteristic function reads
\begin{equation*}
\phi_-(x)=2\left(x^{-x}-1+\exp\left[W\left(1-x^{-x}\right)\right]\right), 
\end{equation*}
again, {after simplifying the expression in \eqref{gen}}, one obtains $G_-(x,1,1)=x^{-x}-1$ and the entropy $S_-$ is restored.

Certainly the Definition \ref{l1} allows to obtain a wide family of entropies with different scaling parameters $(c,d)$ generated by the entropic forms associated to $S_{\pm}$. Albeit, as we are to describe in the remarks below, the classification given by Hanel \& Thurner in \cite{hanel2011} can be directly recovered from \eqref{gen} as well as the particular cases in \eqref{spm}.  

\begin{description}
\item[Remark I] In particular, for $\gamma_\pm(x)=-x \log x+O(-\frac{1}{n!}(x\log x)^n)$, then $W(-\gamma_\pm(x))\to\log x$, and $G_\pm(\cdot)\to g(\cdot)$, where
{
\begin{equation}
\label{hanel}
g(x;c,d) \equiv \sigma\left[\phi(x)-\Gamma[d+1, a(x,c)]e^r\right],
\end{equation}
with $a(x;c)=1-c\log(x)$, $r>0$, hence corresponding to the generalised entropic form given in \cite{hanel2011}}.

\item[Remark II] The universal entropic form \eqref{gen} as for the equivalence class $(c,d)=(1,1)$ yields, $G_\pm(x;1,1)=\pm1\mp x^{\pm x}$, hence the non-parametric entropies $S_\pm=\sum_n^\Omega\left(\pm 1 \mp p_n^{\pm p_n}\right)$ are recovered.

\item[Remark III] The entropic form \eqref{hanel} characterised by the equivalence class $(c,d)=(1,1)$ yields $g(x;1,1)=-x \log x$, restoring the Boltzmann entropy $S_B=-\sum_n^\Omega p_n \log p_n$.
\end{description}

In reference \cite{hanel2011} the authors claim that $g$ as for Eq. \eqref{hanel} does satisfy the asymptotic laws \eqref{law1}-\eqref{law2}. Here we are rather interested in $G_\pm$ characterised by the specific cases to be presented below. In fact, as we will show, the generalised non-extensive entropies $S_\pm$ wholly fulfil the laws \eqref{law1}-\eqref{law2}.

\begin{table*}[t!]
\centering
\begin{tabular}{|r l|c|c|c|}
\hline
\multicolumn{2}{|c|}{Entropy} &  $c$    & $d$ \\ \hline
$p$-entropy $-$ & $S_-(P)=\sum_n^{\Omega}\left(p_n^{-p_n}-1\right)$ 	& 1          & 1  \\ \hline
$p$-entropy $+$ & $S_+(P)=\sum_n^{\Omega}\left(1-p_n^{p_n}\right)$   	& 1          & 1  \\ \hline
W-exponential & $S_{e}(P)=\frac{1-\sum_n^{\Omega}\exp[rW(p_n^{p_n}-1)]}{r-1}$    & $0<r<1$   & 0  \\ \hline
Boltzmann     & $S_B(P)=-\sum_n^{\Omega}p_n\log p_n$    & 1   & 1   \\ \hline
Tsallis       & $S_q(P)=\frac{1-\sum_n^{\Omega}p_n^q}{q-1}$ & $0<q<1$  & 0 \\ \hline
Kaniadakis    & $S_\kappa(P)=-\sum_n^{\Omega}p_n\frac{p_n^\kappa-p_n^{-\kappa}}{2\kappa}$  & $0\leq1-\kappa<1$ & 0   \\ \hline
\end{tabular}
\caption{Generalised entropies of the form \eqref{entropy}. All these are particular cases of \eqref{gen} as for Definition \ref{l1}. Note that $S_\pm(P)$ and $S_B(P)$ are asymptotically equivalent, they correspond to the class $(c,d)=(1,1)$. 
}
\label{t:entropies}
\end{table*}

{We now point out that non-extensive entropies $S_\pm$ guarantee statistics of physical applicability, provided these measures are stable in the sense of Lesche \cite{lesche}. The idea behind is that for a given $\epsilon>0$ there exists a $\delta_\epsilon>0$ such that for two distributions of probability $P$ and $P'$ defined over the number of states $\Omega$ it follows $\|P-P'\|_1<\delta_\epsilon$, thus implying that $\vert S(P)-S(P')\vert<\epsilon S^*(\Omega)$, where $S^*$ is the maximum of entropy, hence it is said that $S$ is stable. In the case of $S_\pm$ the test of stability is formally discussed as follows.

Let $A_\pm(p;t)$ be defined by
\begin{equation}
\label{Apt}
A_\pm(P;t) = \sum_n^\Omega \left(p_n-e_\pm^{-t}\right)\theta\left(p_n-e_\pm^{-t}\right),
\end{equation}
where $\theta(x)$ is the Heaviside theta function and the stretched exponential functions $e_\pm$ are such that $\log_\pm(e_\pm^x)=e_\pm^{\log_\pm(x)}=x$, with $\log_\pm(x) = (\pm 1 \mp x^{\pm x})/x$.

Since the inequality $\vert x\theta(x) - y\theta(y) \vert \leq\vert x-y\vert$ holds, it follows from \eqref{Apt} that $\vert A_\pm(P;t) - A_\pm(P';t) \vert \leq \left\| P-P' \right\|_1$. As well, the entropies $S_\pm$ are now expressed in terms of \eqref{Apt} as
\begin{equation}
S_\pm(P)=\phi +\int\limits_0^\infty\mathrm{d}t\,[1-A_\pm(P;t)],
\end{equation} 
hence we have
\begin{equation}
\begin{split}
\left\vert S_\pm(P)-S_\pm(P') \right\vert &= \left\vert \int\limits_0^\infty \mathrm{d}t [A_\pm(P;t)-A_\pm(P';t)] \right\vert\\
&\leq \int\limits_0^{a_\pm+\log_\pm\Omega}\mathrm{d}t\, \left\vert A_\pm(P;t)-A_\pm(P';t) \right\vert\\
&+\int\limits_{a_\pm+\log_\pm\Omega}^\infty\mathrm{d}t\, \left\vert A_\pm(P;t)-A_\pm(P';t) \right\vert,
\end{split}
\end{equation}
where $a_\pm\geq-\log_\pm\Omega$. In particular, if $a_\pm\geq0$ we can easily compute the integral in the first term, whereas the integral in the second term can be performed by using $(1-e_\pm^{-P+\log_\pm\Omega})\leq A_\pm(P;t)<1$, then we obtain
\begin{equation}
\label{statement}
\left\vert S_\pm(P)-S_\pm(P') \right\vert\leq \|P-P'\|_1(a_\pm+\log_\pm\Omega)+e_\pm^{-a_\pm}R_\pm(\Omega),
\end{equation}
here $R_\pm(\Omega)$ are the residual functions from the integration of $e_\pm^{-t}$ w.r.t. $t$ such that $R_\pm(\infty)\sim1$. Moreover, given that $a_\pm\geq0$, the right-hand side becomes minimum at $a_\pm=-\log_\pm\|P-P'\|_1$, where $\|P-P\|_1<1$, therefore from \eqref{statement} one gets
%
\begin{equation}
\label{proof}
\frac{\left\vert S_\pm(P)-S_\pm(P')\right\vert}{\log_\pm\Omega} \leq  \delta \left(1+\frac{R_\pm(\Omega)}{\log_\pm\Omega} \right) - \delta\log_\pm\delta,
\end{equation}
this follows from the fact that $-x\log_\pm x$ is a nonnegative, continuous function in the interval $[0,1/e_\pm]$, in consequence $\|P-P'\|_1<\delta<1/e_\pm$ and one can conclude that there is an appropriate $\delta_\epsilon$ for every $\epsilon$ such that the right-hand side in \eqref{proof} is a continuous function approaching 0 as $\delta\to0$. In this regard, the values of a given observable represented on the basis of $S_\pm$ statistics should change gently if the state in consideration becomes different to a small amount.
}

On account of the stability criterion as well as those arguments of super-statistics we suggest that any reasonable generalisation to \eqref{gen} depending on $p\log p$ should nearly coincide with the features already considered in $S_\pm$.

Furthermore, the Definition \ref{l1} guarantees that a number of entropies can be derived from \eqref{gen} based on the generators $\gamma_\pm$. In Table \ref{t:entropies} some of the entropies that are straight derivable from \eqref{generators} and \eqref{gen} are portrayed. Among these entropies, the W-exponential entropy is a particular result of this study. At some extent, it can be interpreted as a generalisation of Tsallis entropy \cite{tsallis}, which we discuss in the next section. 

{Further generalised entropies such as the Anteneodo-Plastino measure \cite{ante} can also be classified following the innovative scheme in \cite{korbel}, which constitutes a generalisation to the approach in \cite{hanel2011}. We noted that considering higher-order exponents in accordance with \cite{korbel}, will lead to the same results as for BG entropy and $S_\pm$. Yet our construction based on the generators $\gamma_\pm$ does not generalise \cite{hanel2011} with regard to other possible asymptotic exponents, but with respect to the proposed entropic forms ---  allowing to derive a number of entropies, including $S_\pm$ (see also Table \ref{t:entropies}).}

Aside from $S_\pm$, the functional generators \eqref{generators} do constitute a basis for further entropy measures such as the linear combination $S_0 \equiv \frac{1}{2}( S_+ + S_-) = \frac{1}{2}\sum_n^\Omega(p_n^{-p_n}-p_n^{p_n})$. Likewise, it can be proved that $S_0$ tends asymptotically to the class $(c,d)=(1,1)$, a result to be expected since both $S_+$ and $S_-$ belong to this class as stated by Remark II. {Alternatively, this very fact can be directly seen from the series expansion $S_0=-\sum_n^\Omega\sum_k^\infty \frac{1}{k!}(p_n\log p_n)^k=\sum_n^\Omega\sum_k^\infty\frac{1}{k!}s^k(p_n)$ for $n$ odd. Then, by summing over $k$, one sees that $S_0$ converges to $\sum_n^\Omega\sinh s(p_n)$. Notice that the requirements for thermodynamical stability are covered term by term in the series, to say $s^{k-1}(p_n)>s^k(p_n)$ provided $p\in[0,1]$, thus resembling BG in the limit $\Omega\to\infty$ (Remark III).}

\begin{figure}[h!]
\includegraphics[width=0.4\textwidth]{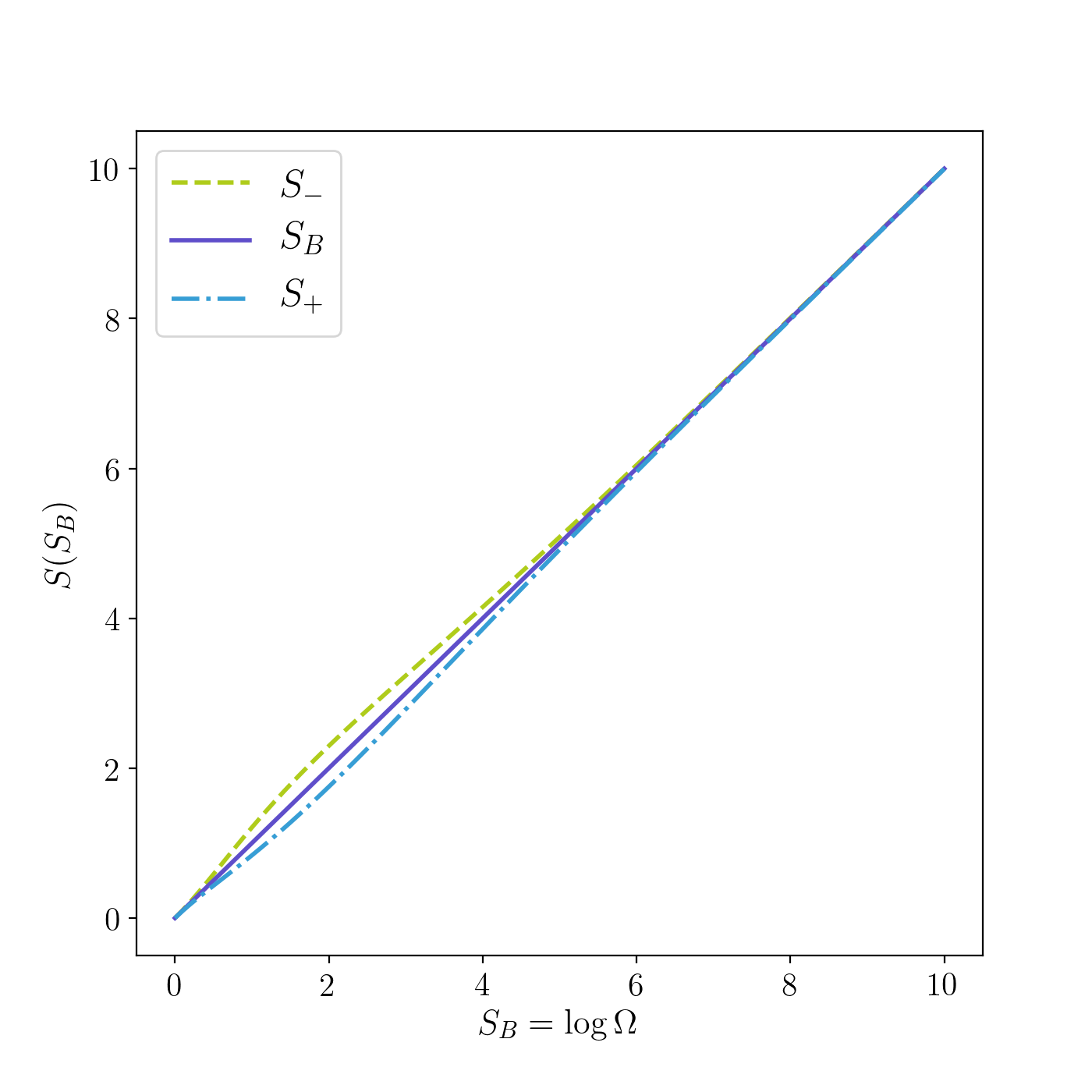}
\caption{Entropies $S_\pm$ as a function of $S_B$ for a uniform distribution.}
\label{f:splus}
\end{figure}

However, unlike entropy $S_B$, the non-extensive $S_\pm$ are suitable to describe a system of few particles beyond the equilibrium \cite{gilvi}. Whether this correlation is weak or strong, is a matter to be explored elsewhere. At this moment we would like to focus our attention on examining a microcanonical configuration $(E,V,N)$ for which the functionals $S_\pm$ attain their maximum at $p_n=1/\Omega$ for every $n$, provided SK2 is satisfied. In the light of this, we get
\begin{equation*}
S_\pm=\pm\Omega\mp\Omega^{1\mp1/\Omega},
\end{equation*}
taking a series expansion, there yields
\begin{equation*}
S_\pm= \sum_{n=1}^\infty (\mp1)^{n+1} \frac{\log^n\Omega}{n!\Omega^{n-1}},
\end{equation*}
the first term in the series corresponds to BG, nonetheless note that the contribution due to the remaining terms tends to be negligible as $\Omega$ grows since $\log^n \Omega < \Omega^{n-1}$, for $n>1$; hence the quotient $\log^n\Omega/\Omega^{n-1}\sim0$ for $\Omega\gg1$. Furthermore, since $S_B=\log\Omega$ as for an equipartition configuration, we get 
\begin{equation}
S_\pm= \sum_{n=1}^\infty (\mp1)^{n+1} \frac{S_B^n}{n!\exp[(n-1)S_B]},
\end{equation}
thus enabling to relate BG to the non-extensive entropies $S_\pm$. This fact has conducted us to the graphs shown in Figure \ref{f:splus} where, as seen, the entropies $S_\pm$ would take into account subtle differences for a statistical system composed of few micro-states. For instance, below 400 ($\log400\approx6$) the differences between $S_{\pm}$ and $S_B$ approaches 0.75\%. 

{
\subsection{A simple toy model}

To serve as an example, we now present a model of diffusion derived from entropies $S_\pm$. A deeper discussion than the presented below is planned to be published elsewhere.
 
We are to obtain a pair of differential equations that reflect anomalous diffusion, namely the case due to $S_+$ ($S_-$) describes  super-diffusion (sub-diffusion). These equations are of the form 
\begin{equation}
\label{diffusion}
\partial_t p_\pm = D \nabla^2 F_\pm[p_\pm], \quad p_\pm=p_\pm(\mathbf{x},t),
\end{equation}
where $D$ is the diffusion coefficient, $p_\pm$ are the distributions of probability in each case and $F_\pm[p_\pm]$ are the effective densities of probability defined as \cite{czegel}
\begin{equation}
\label{formula}
F_\pm[p_\pm]=-\beta\int_0^{p_\pm}\mathrm{d}x\,x\partial_{x}\Lambda_\pm(x),
\end{equation}
with $\Lambda_\pm(x)$ the correspondent generalised logarithms.

To proceed, we first optimise the functional of the form
\begin{equation}
\label{functional}
\Phi = G_\pm(x) - \alpha x - \beta U f_\pm(x),
\end{equation}
here $\alpha$ and $\beta$ are Lagrange multipliers associated with the normalisation condition and conservation of energy $U$. The entropic forms $G_\pm$, in our case, are given by Eq. \eqref{gen}, and the respective weight functions are $f_\pm(x)=x^{x\pm1}$. Differentiating \eqref{functional} w.r.t. $x$, equating to zero and solving for $U$, we can identify the generalised logarithm $\Lambda(x) \equiv U$. For example, in the case of $S_+$ one gets
\begin{equation}
\Lambda_+(x)=\frac{1-x^{-x}+\log x}{1+ x+x\log x},
\end{equation}
similarly for $S_-$.
At this point, we are in the position to compute the effective density related to $\Lambda_+(x)$ via the formula \eqref{formula}. Taking a series expansion, we get 
\begin{equation}
F_+[p_+] = p_++\frac{p_+^2}{4}+\frac{p_+^3}{27}+\frac{p_+^4}{128}+\cdots,
\end{equation}
please note that, after the effective density has been substituted into \eqref{diffusion}, the first term in the series corresponds to the usual diffusion equation --- which in turn would be obtained if BG entropy would have characterised Eq. \eqref{diffusion}. Whereas the presence of nonlinear terms can be due to the appearance of effective, repulsive forces \cite{gilvi}, with a non-negligible participation for systems of high probabilities, that is for those systems with few microstates. 

There are, however, additional physical consequences portrayed by statistics based on $S_\pm$, a detailed perspective on the subject is offered in \cite{ogv,obregon15,Obregon18,CaboObregon18}.
}
%
%
%
%
\section{Further entropic forms}

{
For a self-contained discussion, we are now to review some generalised entropies possessing a composed form $S=F\left(\sum_nG(p_n)\right)$. In this family we find, for example, the Sharma-Mittal \cite{sharma} and R\'enyi \cite{renyi} entropies. However, Tsallis entropy \cite{tsallis}, which belongs to the family \eqref{entropy}, can be recovered from them under specific assumptions. Considering all these entropies, in what follows we are to compute their leading behaviour for a microcanonical configuration as well as their scaling properties in the asymptotic limit.
}

We begin with the Sharma-Mittal entropy, which is non-extensive, this reads
\begin{equation}
\label{sharma}
S_{q,r}\equiv\frac{1}{r-1}\left[1-\left(\sum p_n^q\right)^\frac{1-r}{1-q}\right], \qquad q,r\in\mathbb{R},
\end{equation}
we note that Boltzmann and Tsallis entropies are recovered in the limits $(q,r)\to1$ and $r \to q$, respectively.

Although entropy \eqref{sharma} is not of the form \eqref{entropy}, we are still entitled to classify $S_{q,r}$ by applying directly the asymptotic laws \eqref{law1} and \eqref{law2} to the form $S=(1-G^\frac{1-r}{1-q})/(r-1)$, with $G=\sum_n p_n^q$. A straightforward calculation let us assert that \eqref{sharma} belongs to the class $(c,d)=(r,\infty)$. Noting in particular that due to $d=\infty$, then $S_{q,r}$ will only satisfy SK1-SK3 for $0<r<1$, yet in general $S_{q,r}$ fails to be thermodynamical stable.


{
An interesting result is obtained by expanding $S_{q,r}$ in powers of $r$

\begin{equation*}
-\frac{\log G}{q-1}-\frac{(r-1) \log
   ^2G}{2 (q-1)^2}-\frac{(r-1)^2 \log
   ^3G}{6 (q-1)^3}+O\left((r-1)^3\right),
\end{equation*}
limiting our attention to the special case $r\to1$, there yields the R\'enyi entropy
\begin{equation}
\label{renyi}
S_\alpha \equiv-\frac{\log\sum_n^\Omega p_n^\alpha}{\alpha-1},\end{equation}
where $\alpha$ is the parameter associated with the degree of convexity, which shall be carefully selected for in accordance with \cite{hanel2011} values out from $0<\alpha\leq1$ could compromise the axioms SK2-SK3. Yet, in some applications \cite{korepin} values $\alpha>1$ address to specific purposes.} This entropy is additive and belongs to the class $(1,1)$, one can verify it by a direct application of \eqref{law1} and \eqref{law2}, or alternatively by maximising $S_\alpha$ subject to the normalisation constraint $\sum p_n=1$, hence obtaining $S_B \equiv S_\alpha=\log \Omega$. In other words $S_\alpha$ belongs to the class (1,1) as well.

 

Another interesting fact becomes visible by expanding the R\'enyi entropy as 
\begin{eqnarray}
\label{tsallis}
S_r&=&-\frac{\log\sum p_n^r}{r-1}= -\frac{\sum p_n^r-1}{r-1} + \frac{(\sum p_n^r-1)^2}{2(r-1)}  +\cdots \nonumber\\
&\underset{r\to q}{\approx}&\frac{\sum p_n^q-1}{1-q} \equiv S_q,
\end{eqnarray}
thus obtaining the Tsallis entropy iff $\left\vert\sum p_n^q-1\right\vert\ll1$. There is a further correspondence: A maximisation of Tsallis or R\'enyi entropy, subject to the same constraints but having redefined the Lagrange multipliers, leads to coincident probability distributions.



{The Tsallis entropy belongs to the equivalence class $(c,d)=(q,0)$. Similar to the case of R\'enyi, it would be desired that $0<q\leq1$ in order to avoid violations to SK2-SK3 \cite{hanel2011}. There are, however, physical systems concerning turbulences that exhibit non-BG distributions, which are successfully described by $q$'s far away from unity \cite{Boghosian}.}

As seen, unlike $S_\pm$, it is clear that entropy measure $S_q$ does not fulfil SK4 asymptotically and hence it cannot resemble BG in the thermodynamical limit. 

The rest of the SK axioms are satisfied by $S_q$ as long as $0<q<1$. The maximum of this entropy when subject to the constraint $\sum p_n=1$, is reached at $p_n=1/\Omega$ for all $n$. As a consequence, for an ($E,V,N$) configuration we have
\begin{equation}
\label{tsallis_micro}
S_q=\frac{\Omega^{1-q}-1}{1-q}=\frac{\exp[(1-q)S_B]-1}{1-q},
\end{equation}
where $S_q$ has been expressed as a function of the Boltzmann entropy $S_B$. Besides, one can be aware that \eqref{tsallis_micro} is bounded by the number $\Omega$ of states, which means that $[1+(1-q)S_q]^\frac{1}{1-q}$ is an integer as for a microcanonical ensemble. Using this notation clearly $\lim_{q\to1}S_q=S_B$, yet, it is indeed that for small numbers $\epsilon=\vert q-1\vert$ meaningful deflections of $S_B$ may arise as seen from the expansion
\begin{equation}
S_q=\sum_{n=0}^\infty(-1)^n \frac{(q-1)^n}{(n+1)!}S_B^{n+1}.\label{Sqexp}
\end{equation}

\begin{figure}[h!]
\includegraphics[width=0.4\textwidth]{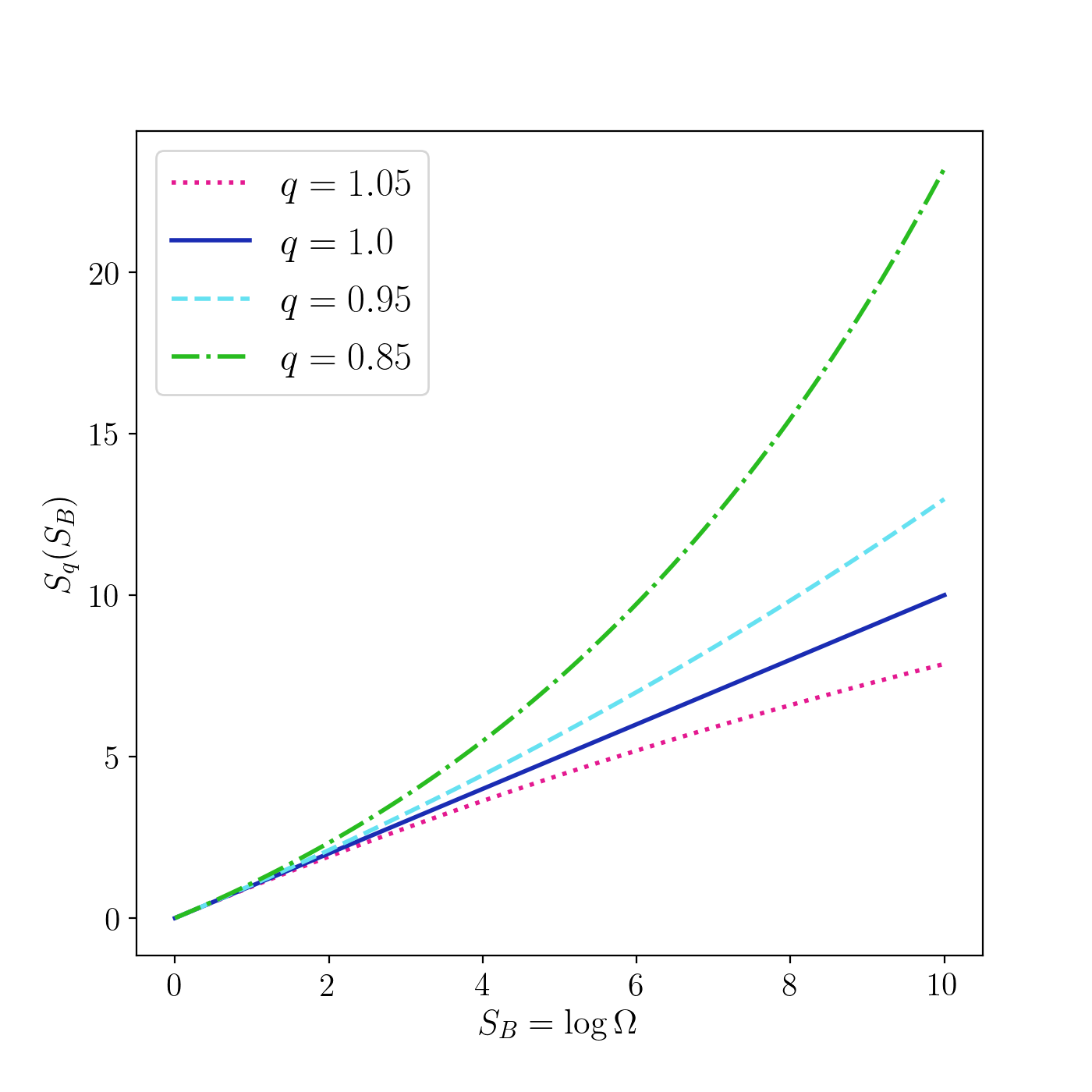}
\caption{Tsallis entropy, $S_q(S_B)$, maximised for a microcanonical ensemble.}
\label{f:tsallis}
\end{figure}

To end this section, let us point out that even for enough small values $\epsilon$ such as $q$ is near the unity, $S_q$ will differ from $S_B$ as the number $\Omega$ of states grows, which can be observed in Figure \ref{f:tsallis}. This is in a way opposite to $S_\pm$ as illustrated in Figure \ref{f:splus} where, for $\Omega\gtrsim400$ and far beyond, these entropies coincide asymptotically with $S_B$; result that would be expected for a large number of microstates in the case of equilibrium or even for a slightly deviation from it. However, one must recall that $S_\pm$ and $S_q$ are constructed based on non-equilibrium assumptions \cite{beck,obregon10}, being these proposals essentially different from $S_B$ in certain $\Omega$ regions. In fact, one could engineer to have $S_q$ near to BG statistics, albeit to achieve that there is a compromise between $\epsilon$ and how large $S_B$ can be. Even for $q$ close to unity, the leading contribution to $S_q$ is given by BG statistics only with a suitable interplay between $\epsilon=\vert q-1\vert$ and $S_{B}$, namely $S_q-S_B=- \sum_{n,i} \frac{(q-1)^{n}}{(n+1)!}p_i \ln p_i^{n+1}$. Hence it is required that $\sum_i p_i \ln p_i^{n+1}\to \epsilon^{k-n}$, where $k>0$. Yet, this is a fictional way of transitioning to the BG statistics limit since it is not possible that such process holds practically. 

\section{Conclusions}

Having surveyed the asymptotic analysis proposed in \cite{hanel2011} we introduced a generalised entropic form in view of the generators \eqref{generators} and, as a consequence, in terms of the fundamental entropies $S_\pm$. As shown, these entropies belong to the same asymptotic class that BG, which means that they do share the same properties in the thermodynamical limit, whereas for a small number of micro-states $S_\pm$ behave in a non-additive way.

It follows that on the basis of entropies $S_\pm$ it is possible to construct a generalised classification of entropy measures. This family will lead to either extensive or non-extensive statistics. As discussed, Tsallis entropy belongs to the generalised entropic forms contained in \eqref{hanel} and therefore in \eqref{gen}. Nonetheless, this entropy does not resembles BG according to the laws \eqref{law1} and \eqref{law2} in the thermodynamical limit as seen from the scaling exponents or, alternatively, by studying the leading contributions as for a microcanonical configuration pointed out in Eq. \eqref{Sqexp} (see also Fig. \ref{f:tsallis}).

Other possible generalisations to BG excluded from \eqref{gen} have been analysed. Such is the case of Sharma-Mittal entropy, having as particular cases: Boltzmann, Tsallis and R\'enyi entropies. In particular, it is worth to note that, both Boltzmann and R\'enyi measures are additive and as a result belong to the class (1,1). Nevertheless, R\'enyi entropy is not of the form \eqref{entropy}. To circumvent this peculiarity, an approximation to Tsallis entropy can be carried out, keeping in mind that in such  limit, R\'enyi's additive property would be not be preserved anymore. This is an interesting example where it is clear that Tsallis and R\'enyi entropies may share certain attributes although not necessarily furnish the same picture.
%
%
%
%
\acknowledgments
J.F. would like to thank the financial support granted by CONACYT (Mexico). O.O. thanks the support of CONACYT Project 257919, UG Project CIIC 188/2019 and PRODEP. N.C.B. thanks the support of CONACYT Project  A1-S-37752 and UG Project CIIC 181/2019.
%
%
%
%
\bibliographystyle{unsrt}
\bibliography{refs}
\end{document}